\documentclass[conference]{IEEEtran}
\IEEEoverridecommandlockouts
\usepackage{cite}
\usepackage{amsmath,amssymb,amsfonts}
\usepackage{algorithmic}
\usepackage{graphicx}
\usepackage{textcomp}
\usepackage{xcolor}

\def\BibTeX{{\rm B\kern-.05em{\sc i\kern-.025em b}\kern-.08em
    T\kern-.1667em\lower.7ex\hbox{E}\kern-.125emX}}
\begin{document}

\title{BlockSDN: Towards a High-Performance Blockchain via Software-Defined Cross Networking optimization\\
}

\author{%
\IEEEauthorblockN{1\textsuperscript{st} Wenyang Jia}
\IEEEauthorblockA{\textit{ICNLab, Shenzhen Graduate School} \\
\textit{Peking University}\\
Shenzhen, P.R.China\\
ORCID: 0009-0000-8876-7027}
\and
\IEEEauthorblockN{2\textsuperscript{nd} Jingjing Wang}
\IEEEauthorblockA{\textit{ICNLab, Shenzhen Graduate School} \\
\textit{Peking University}\\
Shenzhen, P.R.China\\
\texttt{jingjing\_wang@pku.edu.cn}}
\and
\IEEEauthorblockN{3\textsuperscript{rd} Ziwei Yan}
\IEEEauthorblockA{\textit{ICNLab, Shenzhen Graduate School} \\
\textit{Peking University}\\
Shenzhen, P.R.China\\
\texttt{yanzw@pku.edu.cn}}
\and
\IEEEauthorblockN{4\textsuperscript{th} Xiangli Peng}
\IEEEauthorblockA{\textit{The Fifth Electronic Research Institute of MIIT} \\
Guangzhou, P.R.China\\
\texttt{xianglipeng@163.com}}
\and
\IEEEauthorblockN{5\textsuperscript{th} Guohui Yuan*}
\IEEEauthorblockA{\textit{ICNLab, Shenzhen Graduate School} \\
\textit{Peking University}\\
Shenzhen, P.R.China\\
\texttt{yuangh@pku.edu.cn}}
}

\maketitle

\begin{abstract}
The scalability of blockchain systems is constrained by inefficient P2P broadcasting, as most existing optimizations focus only on the logical layer without considering physical network conditions. To address this, we propose BlockSDN, the first SDN-based integrated architecture for blockchain. BlockSDN employs a distributed control plane for a global network view, a graph engine for hierarchical clustering, and a hybrid macro–micro neighbor selection with hierarchical broadcasting. A dedicated simulation platform shows that BlockSDN reduces global block synchronization time by 65\% and 55\% compared to Gossip and Mercury, respectively. These results highlight the potential of SDN-enabled cross-layer coordination to significantly enhance blockchain scalability and performance.
\end{abstract}

\begin{IEEEkeywords}
Software Defined Network, Blockchain, Block transmission
\end{IEEEkeywords}

\section{Introduction}

Blockchain, as a revolutionary distributed ledger technology, is transforming industries such as finance, supply chain, and healthcare through its decentralized, transparent, and tamper-resistant features. However, a major obstacle hindering its large-scale deployment lies in performance bottlenecks—particularly limited transaction throughput. This issue stems largely from the inefficient communication mechanisms within the underlying peer-to-peer (P2P) network layer. In a globally distributed blockchain system, newly generated transactions and blocks must be rapidly and reliably propagated to the majority of nodes to reach consensus. Yet, propagation delays remain a serious concern: for example, broadcasting a 1MB Bitcoin block takes approximately 80 seconds to reach 90\% of peers, and Ethereum requires around 10 seconds~\cite{decker2013information}.

These high delays directly restrict block production rates and transaction processing capacity, while also exacerbating consistency and security risks. In competitive mining systems, delayed block propagation increases the likelihood of forks, wasting computational resources on orphaned blocks and undermining system integrity. More importantly, the blockchain P2P network must ensure not only delivery efficiency, as in traditional P2P systems, but also strict global consistency. This demands far more from the broadcast protocol and network topology. Although widely adopted due to its robustness, the Gossip protocol suffers from randomness and high message redundancy, contributing to network congestion and degraded latency~\cite{fanti2018compounding}.

To mitigate these challenges, existing research primarily focuses on two directions: optimizing network topology and improving broadcast protocols. For topology design, methods include geographic clustering~\cite{li2019network}, reputation-based hierarchical structures~\cite{wang2021reputation}, and adaptive peer selection via online learning~\cite{bao2020adaptive}. For broadcast schemes, researchers have proposed tree-based propagation strategies and overlay-based broadcast optimizations~\cite{zhang2020broadcasttree}. However, a fundamental limitation persists across these approaches: they predominantly perform optimization at the logical layer of the blockchain, with limited or no awareness of the real-time status of the underlying physical network (e.g., bandwidth, latency, or topology)~\cite{gencer2018decentralization}.

This decoupling between logical and physical layers gives rise to several unresolved issues:

\begin{itemize}
    \item {Neglect of resource heterogeneity}:Cluster heads or relay nodes are typically chosen by structural metrics (e.g., topology position), overlooking heterogeneity in computing or bandwidth and causing under-provisioned nodes to become bottlenecks.
    \item {Lack of global coordination}:Neighbor selection is often decentralized and heuristic, lacking centralized guidance, which leads to fragmented overlays and suboptimal topologies.

    \item {High cost of dynamic broadcast tree construction}: While broadcast trees mitigate redundancy, their dynamic construction and maintenance incur high computation and communication overhead, limiting practicality under high-throughput workloads.

\end{itemize}

These challenges stem from the traditional TCP/IP architecture, whose decoupled layers and decentralized control hinder global visibility and dynamic optimization for blockchain requirements such as rapid, synchronized block propagation.

To fundamentally address these cross-layer coordination challenges, we introduce the architectural paradigm of \t{Software-Defined Networking (SDN)} into blockchain systems. We propose \textbf{BlockSDN}, a novel SDN-enabled integration framework that bridges the blockchain logical layer with the underlying physical network. SDN separates the control and data planes, enabling a logically centralized controller to acquire a global network view and apply programmable logic~\cite{mckeown2008openflow}. Leveraging this capability, BlockSDN dynamically monitors real-time network conditions and uses this information to optimize both the logical overlay and broadcast mechanisms at the blockchain layer. 

BlockSDN embeds a graph engine in the SDN control plane to unify node geolocation, link states, and logical connections. Leveraging this abstraction, it employs a hybrid coordination mechanism with a macro–micro neighbor selection algorithm and a hierarchical broadcast strategy, achieving global optimization with local adaptivity.

The main contributions of this paper are summarized as follows:

\begin{itemize}
    \item {A novel blockchain–network integration architecture}:We propose BlockSDN, a three-layer SDN-based framework that integrates a graph engine into the control plane for unified modeling of blockchain overlays and physical network states, enabling full-stack cross-layer awareness and coordination.

    \item {Microscopic-Macroscopic Collaborative Data Synchronization}: We propose a comprehensive data synchronization optimization framework comprising three key components: (1) a resource-aware hierarchical partitioning algorithm that performs both clustering and layering of network nodes; (2) a hybrid neighbor selection mechanism that coordinates macroscopic controller-driven global recommendations with microscopic node-level adaptive decisions; and (3) a hierarchical topology-based broadcast algorithm that significantly reduces both propagation latency and redundant transmission overhead.
    \item {Comprehensive Performance Evaluation and Validation}: We conduct extensive experiments across diverse network scales, communication loads, and physical topologies. Results show that BlockSDN significantly outperforms state-of-the-art approaches such as Gossip and Mercury in block dissemination time, throughput, scalability, and topological adaptability.

\end{itemize}

\section{BlockSDN System Design}

\subsection{Architectural Design Goals and Solution Approaches}

Current network architectures fall short of meeting the performance requirements of blockchain systems~\cite{jin2020accelerating}, particularly in terms of throughput and latency. These limitations are reflected in the following aspects:

\subsubsection{Limitations of the Traditional TCP/IP Architecture}
The traditional TCP/IP model, while ensuring reliability, lacks the real-time network awareness and cross-layer coordination required for the rapid, global synchronization of blockchain systems~\cite{ko2020p4}. Its static, protocol-defined logical connections cannot adapt to dynamic physical network conditions, creating a mismatch between logical and physical topologies~\cite{liu2019survey}. This structural deficiency fundamentally restricts block propagation efficiency and limits system scalability.

\subsubsection{Adaptability Limitations of Existing SDN Architectures}
While Software-Defined Networking (SDN) excels at centralized management by decoupling the control and data planes, its architecture is ill-suited for the unique demands of blockchain~\cite{he2020novel}, such as high-frequency broadcasts and global synchronization. In large-scale, high-throughput scenarios, the centralized controller becomes a performance bottleneck, causing response delays and coarse scheduling. Furthermore, its forwarding mechanisms are optimized for physical paths, lacking awareness of blockchain's logical overlays~\cite{khatri2020optimizing}. Consequently, conventional SDN fails to support the high-throughput, low-latency, and dynamic requirements of modern blockchain topologies.

To address the above limitations, we propose \textbf{BlockSDN}, a blockchain–network integrated architecture based on SDN. BlockSDN is designed with the following objectives:

\subsubsection{High Performance}
To address blockchain’s high-throughput and low-latency demands, BlockSDN employs a macro–micro coordinated neighbor selection algorithm, where the controller computes and distributes optimal neighbor sets for low-latency, load-balanced connections. A complementary hierarchical broadcast mechanism suppresses redundancy, accelerating block dissemination and consensus convergence.

\subsubsection{Scalability}
To support scaling blockchain nodes and applications, BlockSDN integrates a graph engine into the SDN controller and adopts clustering-based synchronization for network management. Nodes are dynamically grouped by characteristics with parallel intra-cluster synchronization, while the global view enables dynamic cluster restructuring to maintain stability and throughput as node count grows.

\subsubsection{Adaptivity}
To handle dynamic node changes and load fluctuations, BlockSDN employs an adaptive mechanism driven by fork-rate feedback. Via northbound APIs, it monitors fork rates as a proxy for synchronization health, and upon detecting increases, the controller re-collects network metrics and adjusts neighbor selection strategies.

\subsubsection{Robustness}
BlockSDN mitigates single-point failures through a distributed control-plane architecture, where multiple cooperative controllers synchronize state via control-domain partitioning. In case of failure or overload, remaining controllers seamlessly assume scheduling, ensuring continuous control, fault tolerance, and service reliability in complex environments.

\begin{figure}[t]
  \centering
  \includegraphics[width=0.9\columnwidth,trim=110 20 60 20,clip]{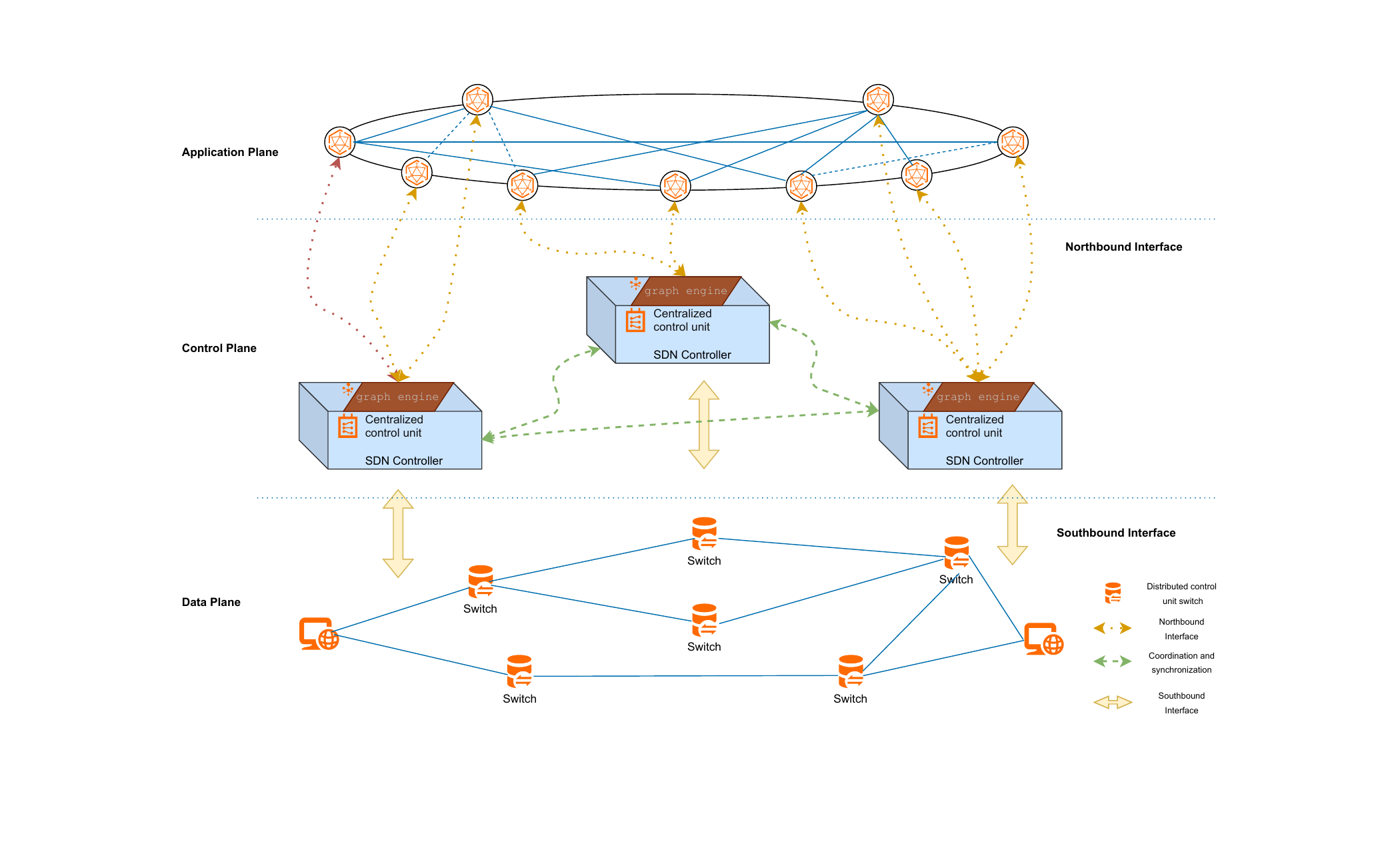}
  \caption{Integrated Architecture of Blockchain Network Based on SDN}
  \label{fig:fig1}
\end{figure}

\subsection{The BlockSDN Architecture}

\subsubsection{Overview}
BlockSDN leverages the SDN control plane bidirectionally: supporting blockchain operation above while managing network infrastructure below. Its primary goal is to enhance network awareness and responsiveness to synchronization tasks, offering tailored optimizations for block dissemination and synchronization to improve efficiency and ensure reliable, high-performance blockchain execution.

The architecture of BlockSDN, illustrated in Figure.~\ref{fig:fig1}, consists of three layers: the {application plane}, the {control plane}, and the {data plane}, each responsible for distinct functionalities.

\subsubsection{Application Plane}
The application plane, consisting of the blockchain network, serves as the core service layer, handling transaction processing and block generation. As a decentralized and open infrastructure typical of public blockchains, it specifies synchronization demands and service requirements for the underlying network.

\subsubsection{Control Plane}
The control plane integrates SDN controllers with an embedded graph engine to deliver global visibility, unified scheduling, and dynamic optimization. By coordinating multiple controllers into a logically centralized yet physically distributed fabric, BlockSDN improves scalability and fault tolerance; with global programmability it refines neighbor selection and overlay topology, monitors fork rates via northbound APIs to trigger reconfiguration, and partitions large networks into clusters for parallel broadcasts—shortening paths, reducing latency, and increasing concurrency.

\subsubsection{Data Plane}
The data plane comprises the physical network—blockchain nodes and forwarding devices—responsible for packet forwarding and local state monitoring. In each control cycle, distributed rule-execution units collect node- and link-level metrics and report them to the control plane, enabling a fine-grained, up-to-date network view that informs real-time neighbor selection.

\subsubsection{Architectural Interaction Design}

\begin{figure*}[t]
    \centering
    \includegraphics[width=1.0\textwidth,trim=110 20 100 20,clip]{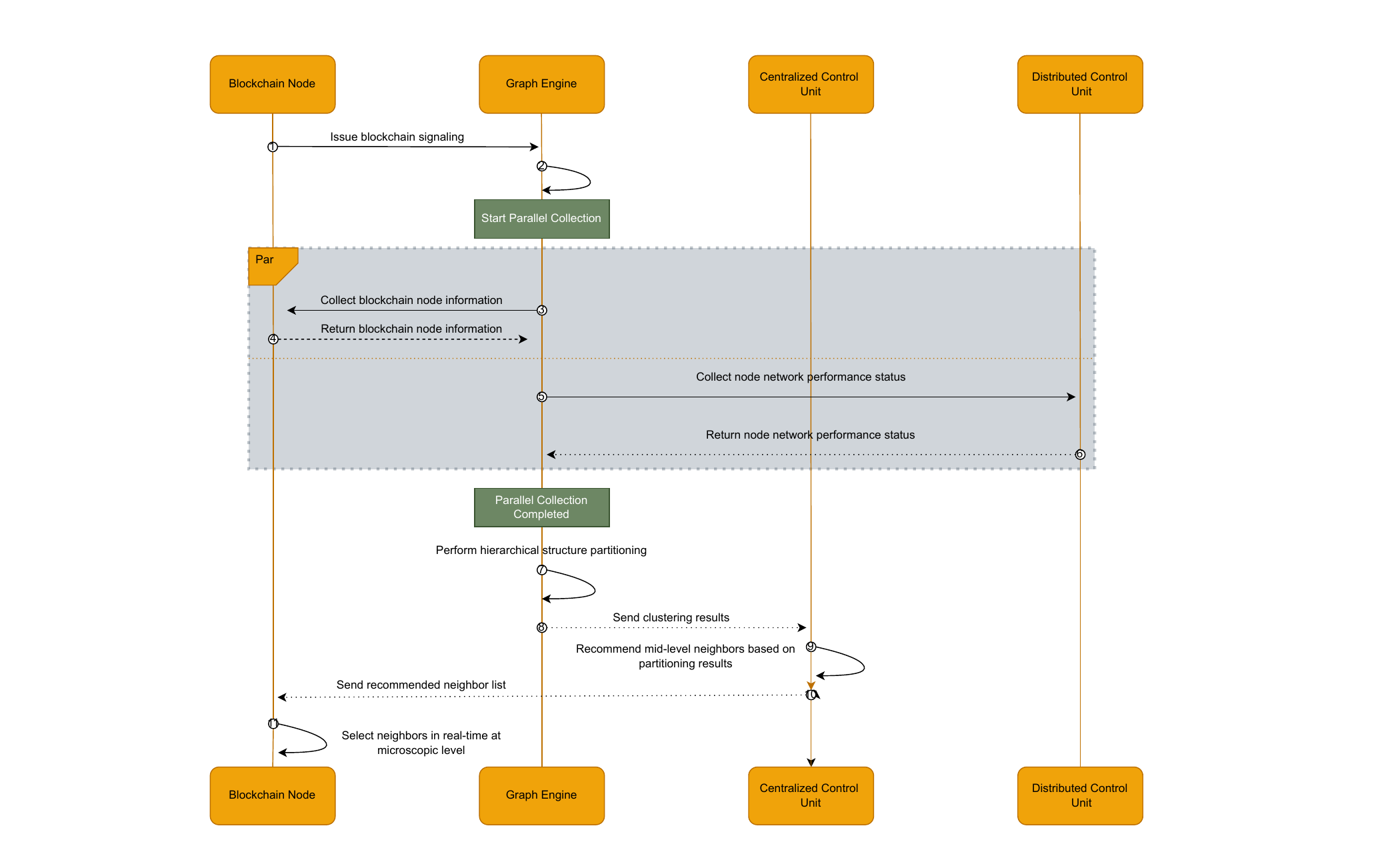}
    \caption{Integrated Architecture of Blockchain Network Based on SDN}
    \label{fig:fig2}
\end{figure*}

Leveraging SDN programmability and the graph engine’s clustering, BlockSDN efficiently supports blockchain synchronization via dynamic topology optimization and rapid responsiveness to upper-layer requirements. Performance, however, depends on both individual component capabilities and their coordinated, dynamic cooperation.

As illustrated in Figure.~\ref{fig:fig2}, the timing diagram presents the system's runtime logic and interaction patterns. The vertical axis represents the time sequence, while the horizontal axis denotes core architectural components—including blockchain nodes (BN), the graph engine (GE), the centralized control unit (CCU), and distributed control units (DCU). Standard timing notations are used to visualize the message-passing order and collaboration procedures. The overall process is divided into two main phases:

{(1) Information Collection and Clustering Phase (Steps 1--7):}  
This phase acquires real-time network state and node attributes from blockchain participants, then applies clustering and hierarchical modeling to produce an accurate, responsive view of network dynamics that enables informed neighbor selection and broadcast-path optimization.

{(2) Macro--Micro Coordinated Neighbor Selection Phase (Steps 8--11):}  
This phase coordinates the controller and blockchain nodes bidirectionally to produce globally guided, locally adaptive neighbor connections—yielding efficient, robust, and latency-aware logical topologies—and, together with the preceding phase, completes BlockSDN’s closed-loop control for topology reconfiguration, enabling downstream broadcast optimization and performance evaluation.

\subsection{Evaluation}
We conducted a systematic experimental evaluation across four dimensions—block propagation time, throughput, scalability, and topology adaptability—under varying network scales, communication loads, and physical topologies to validate BlockSDN's performance.

\subsubsection{Block propagation time}

\begin{figure}[t]
    \centering
    \begin{minipage}[t]{0.42\textwidth}
        \centering
        \includegraphics[width=1\linewidth]{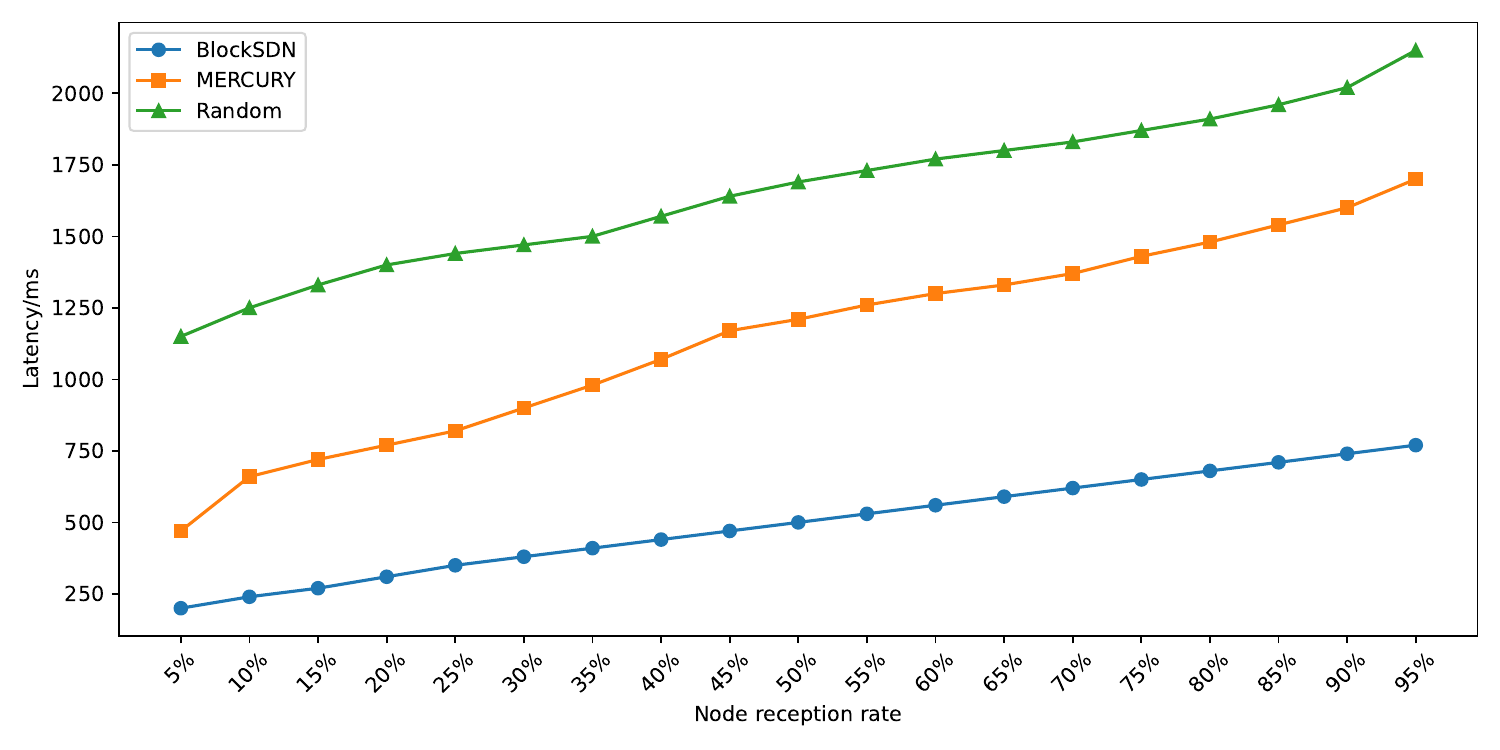}
        \caption{Block propagation time under varying node reception ratios}
        \label{fig:fig51}
    \end{minipage}
    \hfill
    \begin{minipage}[t]{0.42\textwidth}
        \centering
        \includegraphics[width=\linewidth]{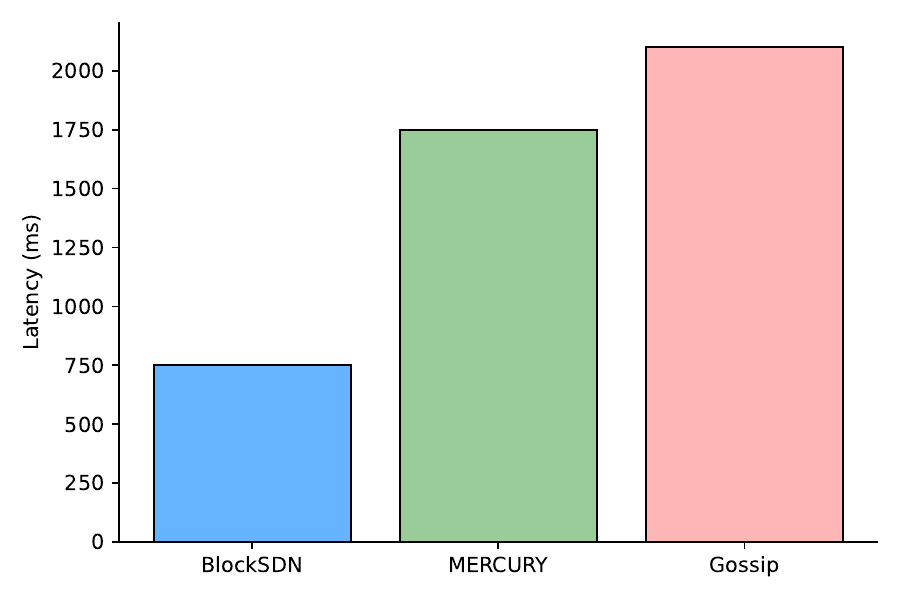}
        \caption{Average delay for transaction propagation across the entire network}
        \label{fig:fig52}
    \end{minipage}
    \hfill
    \begin{minipage}[t]{0.42\textwidth}
        \centering
        \includegraphics[width=\linewidth]{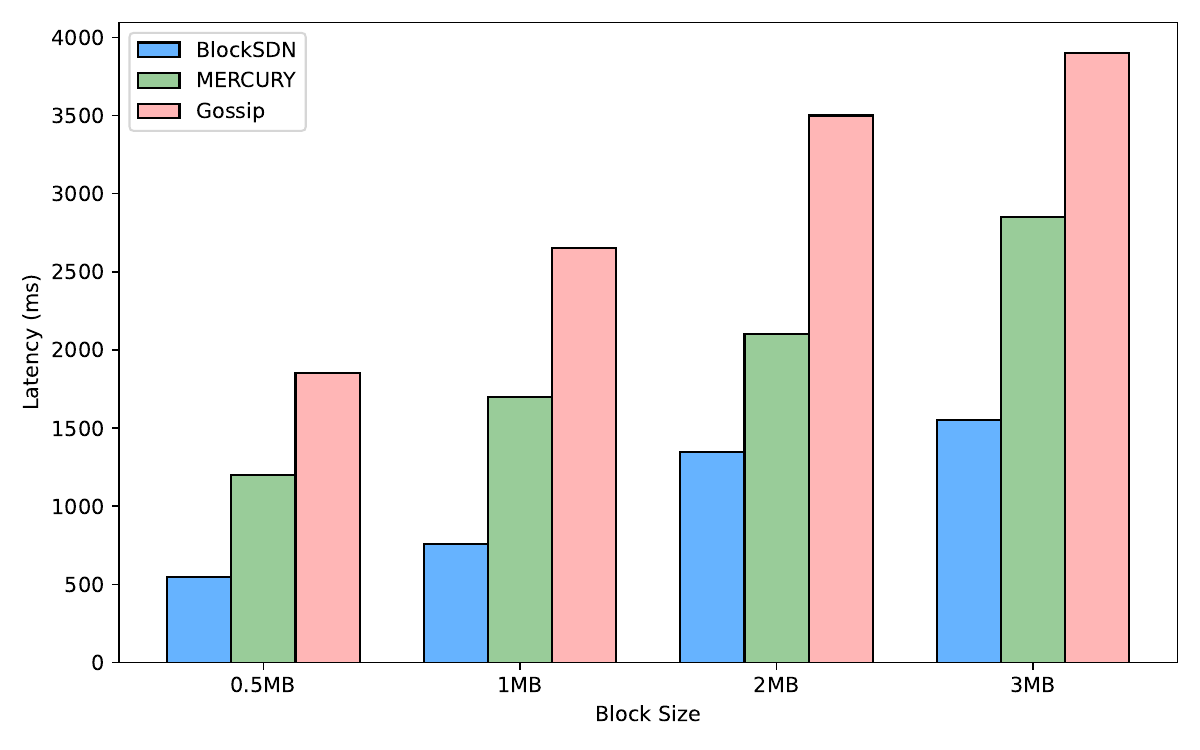}
        \caption{Block transmission time for different block sizes}
        \label{fig:fig53}
    \end{minipage}
\end{figure}

As shown in Figure.~\ref{fig:fig51}, compared with Gossip and Mercury, BlockSDN significantly shortens the block propagation time under all synchronization ratio conditions. For example, when the synchronization ratio reaches 50\%, Gossip takes 1647\,ms, Mercury~\cite{zhou2023mercury} takes 1202\,ms, while BlockSDN takes only 500\,ms, reducing the propagation delay by 58\% and 69\% compared to Mercury and Gossip, respectively. 

Further analysis of the delay for a transaction to reach all nodes is shown in Figure.~\ref{fig:fig52}. Gossip takes 2146\,ms, Mercury takes 1700\,ms, while BlockSDN only 764\,ms, reducing the delay by 55\% and 65\%. This fully demonstrates the significant advantage of BlockSDN in high-concurrency block propagation scenarios. 

Moreover, the propagation time curve of BlockSDN grows more smoothly with increasing synchronization ratio, showing stronger stability. When the synchronization ratio increases from 20\% to 95\%, Mercury requires 1229\,ms, while BlockSDN requires only 565\,ms. In summary, BlockSDN significantly improves broadcast performance.

As shown in Figure.~\ref{fig:fig53}, with block size increasing from 0.5\,MB to 3\,MB, all algorithms exhibit scaling behavior. However, BlockSDN consistently achieves lower delay. For example, when block size is 0.5\,MB, Gossip requires 1864\,ms, Mercury 1188\,ms, and BlockSDN only 548\,ms. At 3\,MB, Gossip takes 3870\,ms, Mercury 2864\,ms, while BlockSDN takes 1545\,ms. Thus, BlockSDN maintains clear advantages under large data volume.

\subsubsection{Network throughput}
\begin{figure}[t]
    \centering
    \begin{minipage}[t]{0.48\textwidth}
        \centering
        \includegraphics[width=\linewidth]{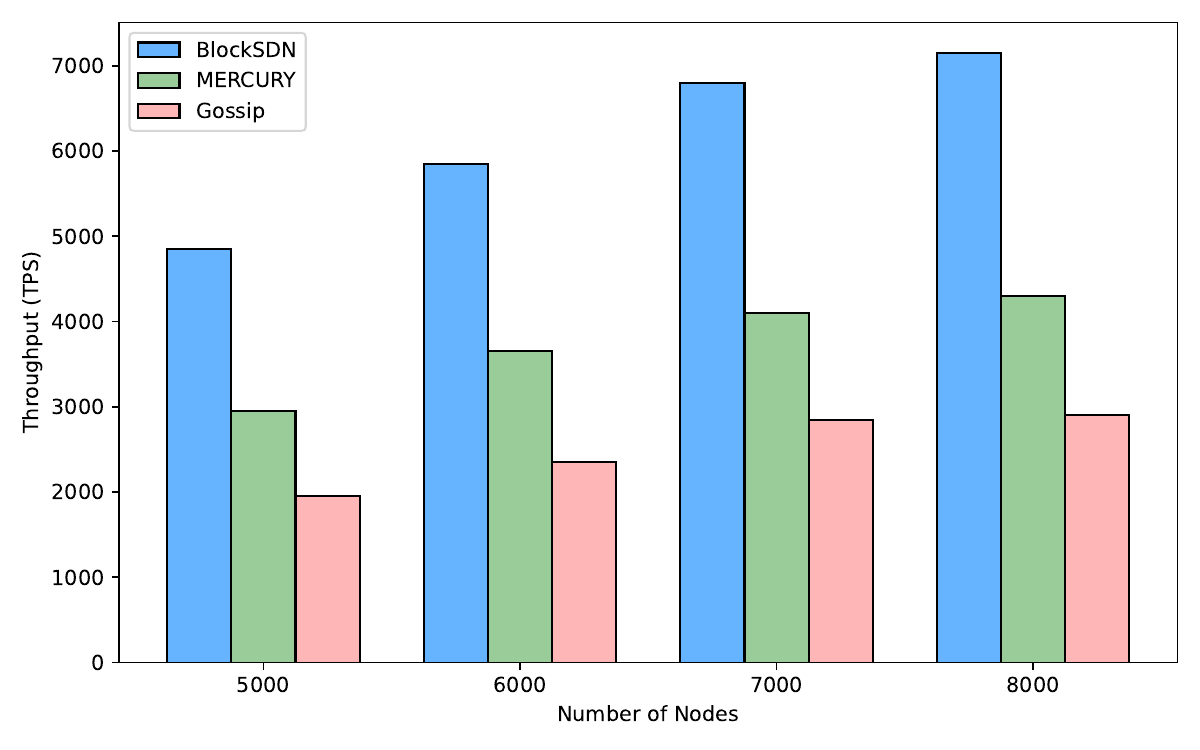}
        \caption{Network throughput under increasing node scale}
        \label{fig:fig54}
    \end{minipage}
    \hfill
    \begin{minipage}[t]{0.45\textwidth}
        \centering
        \includegraphics[width=1.02\linewidth,trim=20 0 10 10,clip]{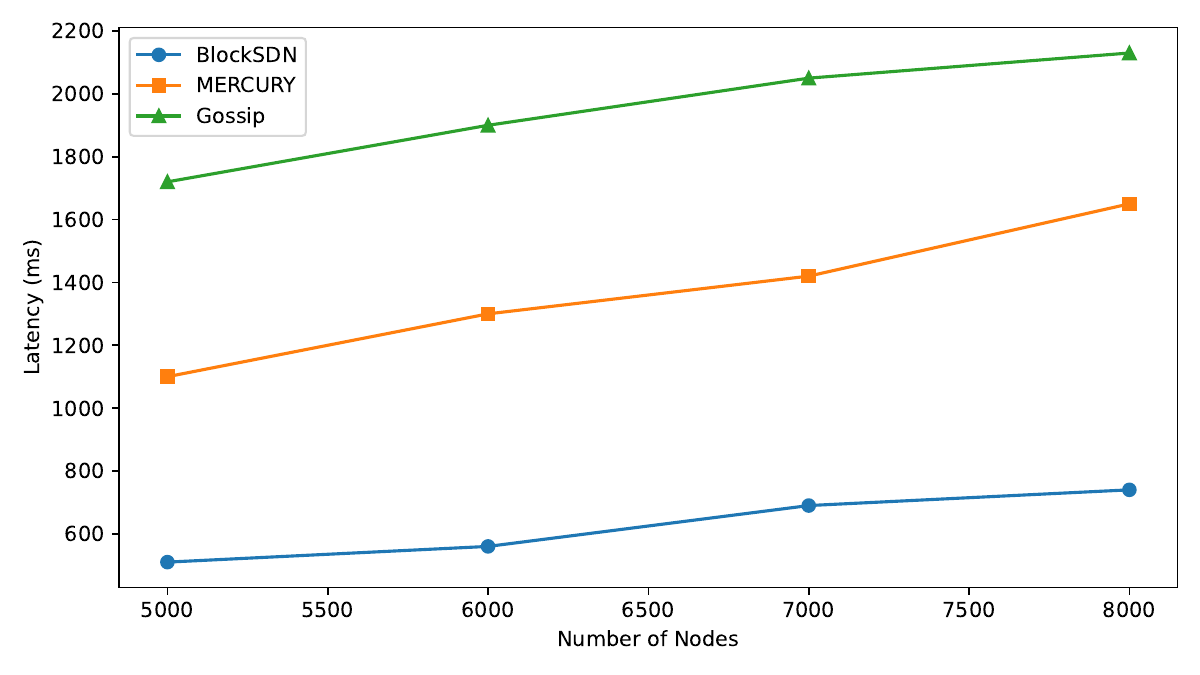}
        \caption{Network scalability with growing network size}
        \label{fig:fig55}
    \end{minipage}
\end{figure}

This part evaluates the throughput of the three strategies, i.e., the number of transactions the system can process per unit time. Results are shown in Figure.~\ref{fig:fig54}. Throughput of all algorithms increases with network scale because more nodes yield more transmitted blocks. 

When the scale increases from 5000 to 8000 nodes, BlockSDN improves by 2283\,TPS, Mercury by 1365\,TPS, and Gossip only by 910\,TPS. However, as the network reaches 8000 nodes, throughput growth slows due to structural complexity and bandwidth limits.

\subsubsection{Network scalability}
This section evaluates scalability by observing block transmission time when the node reception ratio is fixed at 95\% and the network size increases from 5000 to 8000 nodes. Results in Figure.~\ref{fig:fig55} show that all schemes experience increased delay due to higher topology complexity and load. 

However, BlockSDN grows more slowly: Mercury increases by 506\,ms, while BlockSDN increases only 259\,ms (49\% reduction). This indicates that BlockSDN more effectively handles load expansion, maintaining lower latency and higher stability.

\subsubsection{Network topology adaptability}
This section evaluates adaptability of broadcast algorithms to diverse topologies: ring, star, and tree. 

\begin{figure}[t]
    \centering
    \includegraphics[width=1.0\linewidth]{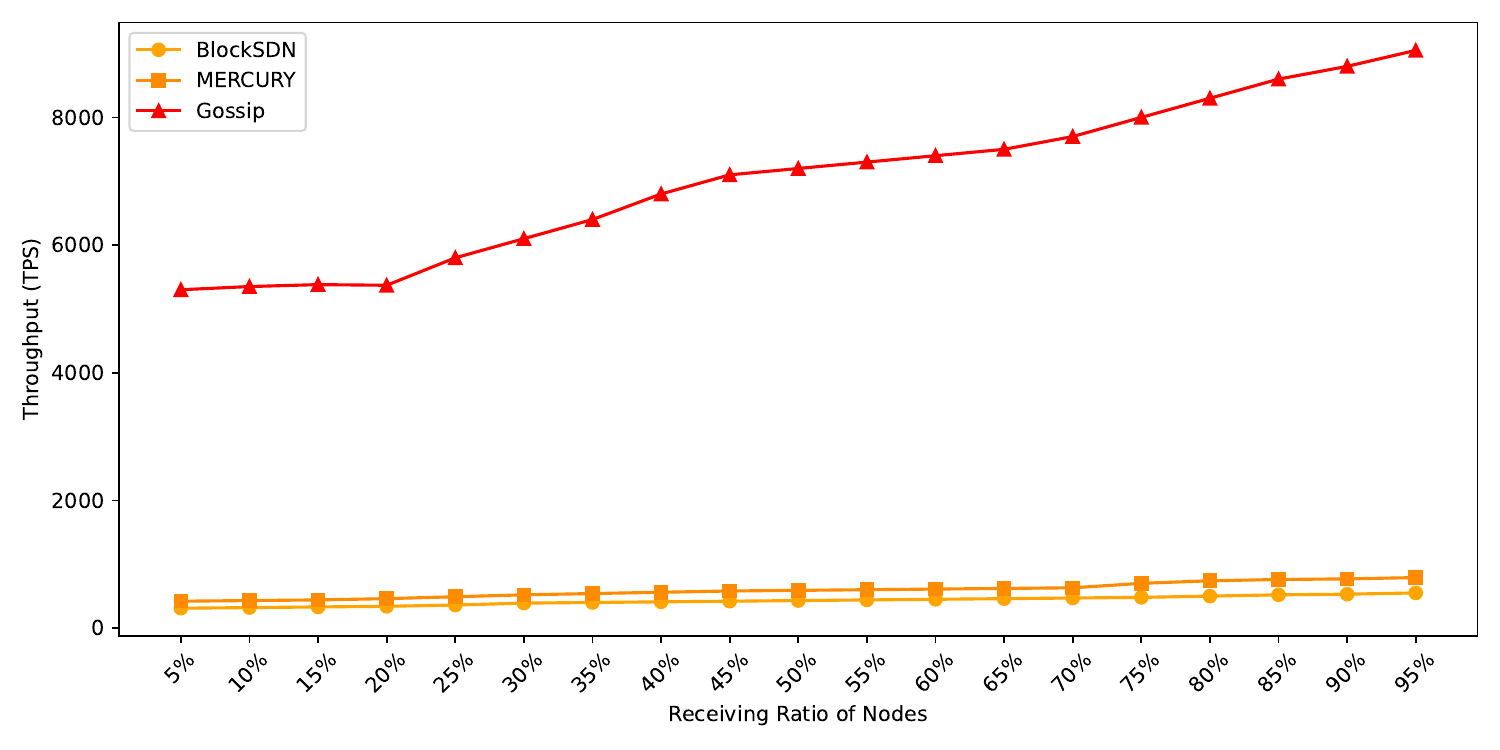}
    \caption{Block propagation time under ring topology}
    \label{fig:fig56}
\end{figure}

In the \textit{ring topology}, with 1000 nodes (10 local rings plus backbone), Gossip requires 9405\,ms, Mercury 949\,ms, and BlockSDN only 560\,ms at 95\% reception. Gossip suffers from random neighbor selection, while BlockSDN and Mercury exploit structure.

\begin{figure}[t]
    \centering
    \includegraphics[width=1.0\linewidth]{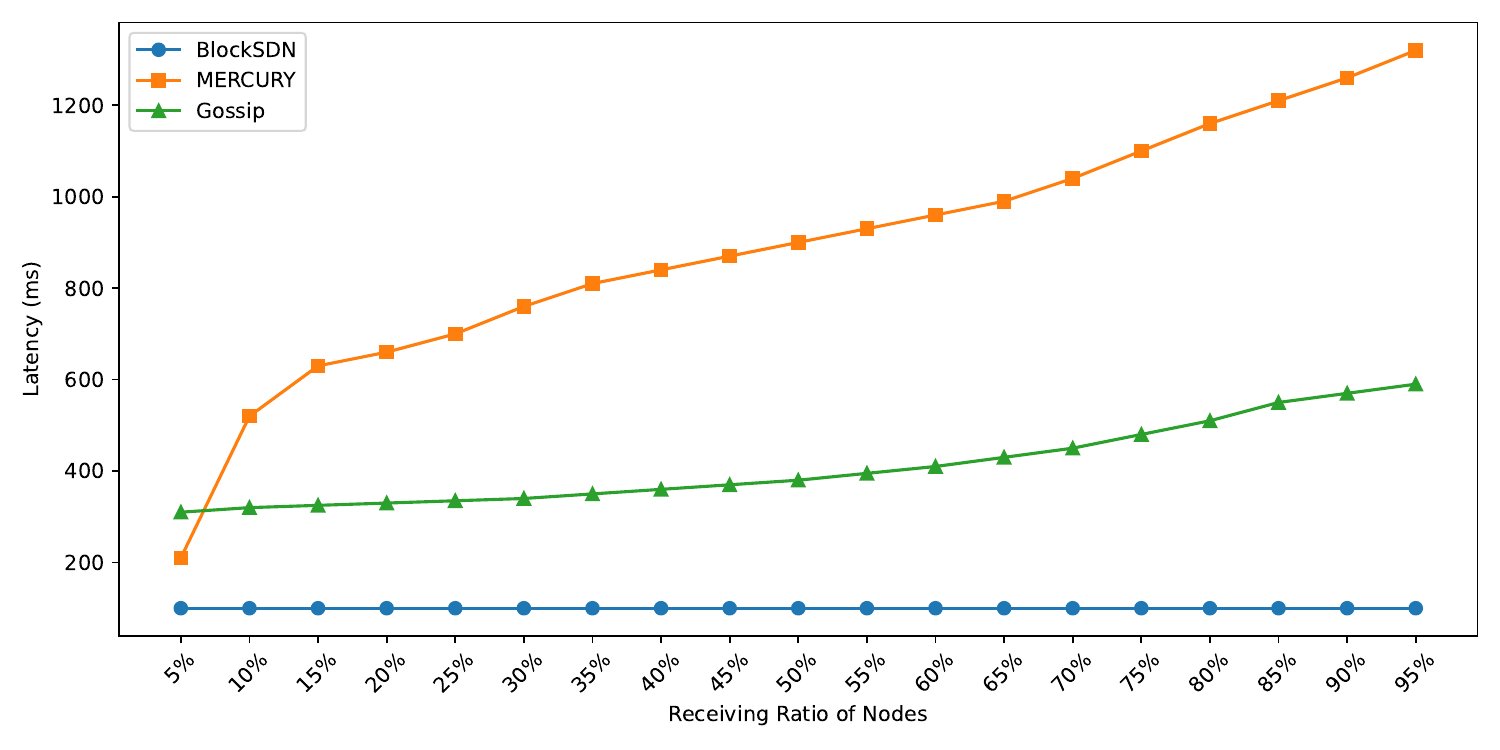}
    \caption{Block transmission time in star topology}
    \label{fig:fig57}
\end{figure}

In the \textit{star topology}, with 1000 nodes, BlockSDN requires only 117\,ms, Gossip 604\,ms, and Mercury 1339\,ms. Gossip benefits from central node bandwidth, while Mercury’s structural optimization becomes ineffective. BlockSDN efficiently manages central nodes, reducing redundancy.

\begin{figure}[t]
    \centering
    \includegraphics[width=1.0\linewidth]{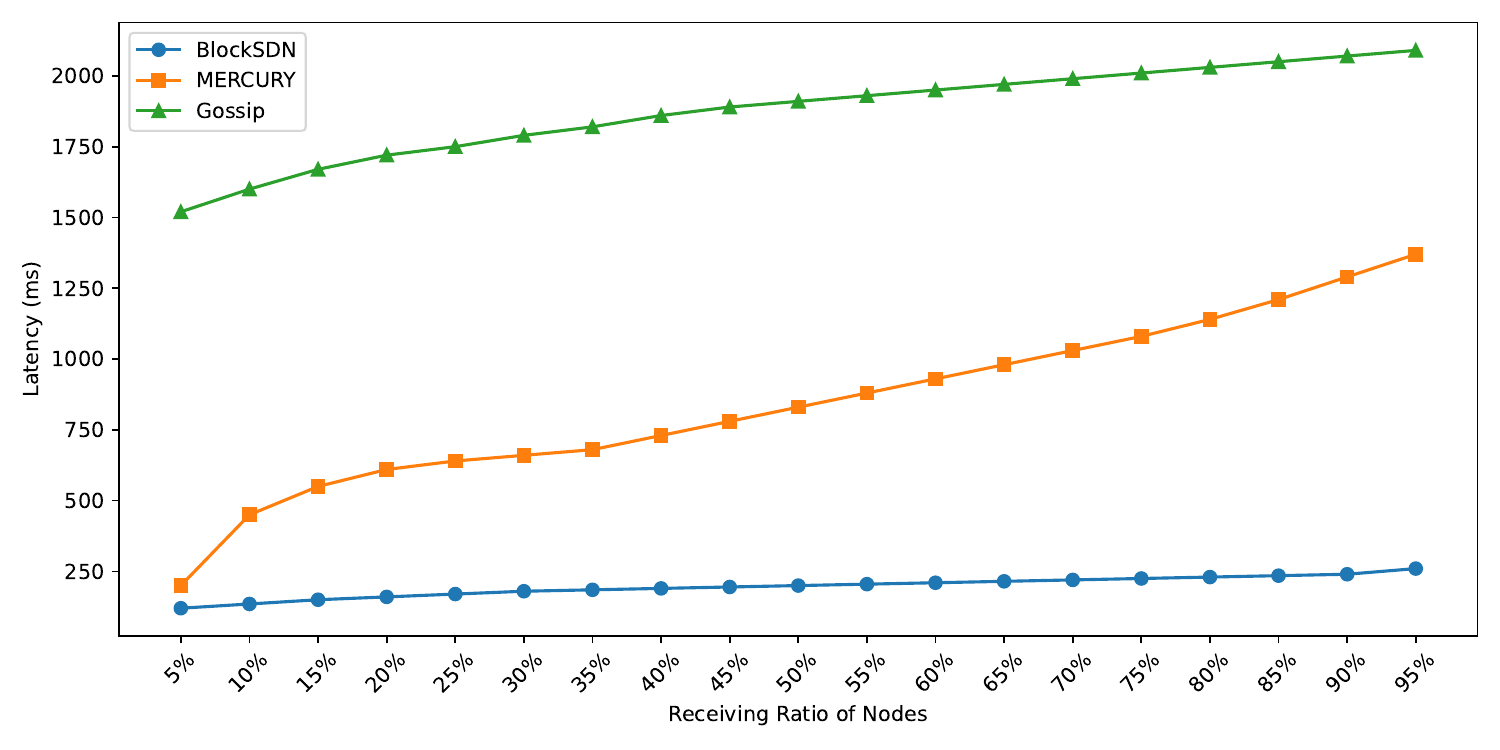}
    \caption{Block transmission time in tree topology}
    \label{figure:figure58}
\end{figure}

In the \textit{tree topology} (1000 nodes, depth five), BlockSDN achieves 260\,ms, Mercury 1356\,ms, and Gossip 2090\,ms. BlockSDN uses hierarchical scheduling to reduce redundant forwarding. 

Across ring, star, and tree topologies, Gossip suffers from redundant paths, Mercury performs inconsistently, and BlockSDN consistently achieves minimal delay. This is attributed to its structured design, global awareness, and adaptive neighbor scheduling, enabling efficient broadcasting under diverse topologies.

\section{Conclusion}
As blockchain networks grow and application scenarios diversify, scalability has become a key bottleneck ~\cite{huang2020ien3}. Purely intra-blockchain optimizations are insufficient; co-optimization with the physical network is required. Yet the rigid, layered TCP/IP architecture lacks visibility into synchronization needs and cannot promptly respond.

SDN decouples control and data planes and offers centralized, programmable control, improving flexibility and global management ~\cite{lei2020ien}. However, little work targets large-scale blockchain synchronization ~\cite{sun2019sparse}. This paper explores SDN-based scalability optimization to enhance network synchronization. The main contributions are:

\begin{itemize}
  \item[(1)] To address the limitations of traditional network architectures in perceiving synchronization demands and the lack of global visibility during data synchronization, we propose a novel SDN-based blockchain network architecture, \textbf{BlockSDN}. Based on traditional SDN, this architecture integrates a graph engine and employs clustering algorithms to partition large-scale blockchain networks into multiple sub-clusters, enhancing the parallelism and efficiency of block propagation.

  \item[(2)] In response to the limitations of local-view-based topology optimization and the inefficiencies and redundancy of the Gossip broadcast protocol in current blockchain systems, we propose a \t{meso-micro coordinated synchronization method} based on the BlockSDN architecture.

\item[(3)] To verify the performance and feasibility of \textbf{BlockSDN}, this paper develops an SDN-based blockchain network simulation tool. The tool adopts a discrete event-driven mechanism to construct a multi-layer architecture and supports modular integration and comparison of various algorithms, including BlockSDN, Mercury, and Gossip. 

This simulation platform models both the propagation process at the blockchain layer and the state evolution of the underlying physical network, thereby realistically reproducing communication behaviors under complex environments. Simulation results demonstrate that, compared to Mercury and Gossip, \textbf{BlockSDN} significantly reduces block propagation latency, lowering the total network-wide broadcasting time by 55\% and 65\% respectively, which validates its superior scalability and stability in large-scale networks.
\end{itemize}

\section*{Acknowledgment}
This work was supported in part by the Shenzhen Key Research Project (JCYJ20220818100810023).

\end{document}